\documentclass[usenatbib,usegraphicx]{mn2e}
\usepackage{times,amsmath,amssymb,afterpage}
\newcommand{\kms}{km s$^{-1}$ }

\begin{document}

\title{Periodicity in Class II methanol masers in high mass star forming regions}

\author[Goedhart et al.]{
S. Goedhart\thanks{E-mail: sharmila@ska.ac.za}$^{1,2,3}$, J. P.  Maswanganye$^{2,3}$, M. J. Gaylard$^{2}$ and D. J. van der Walt$^{3}$  \\
$^1$SKA SA, 3rd Floor, The Park, Park Rd, Pinelands, 7405, South Africa\\
$^2$Hartebeesthoek Radio Astronomy Observatory, PO Box 443, Krugersdorp, 1740, South Africa \\
$^3$School of Physics, North-West University, Potchefstroom campus, Private Bag X6001, Potchefstroom, 2520, South Africa\\
}

\date{Accepted ;      Received ;      in original form }

\pagerange{\pageref{firstpage}--\pageref{lastpage}}
\pubyear{2012}

\maketitle

\begin{abstract}

We report the results of 10 years of monitoring of six regularly varying 6.7 GHz methanol masers using the Hartebeesthoek 26m telescope. Observations were done at intervals of 1--2 weeks, with faster sampling during flaring episodes. Four of the sources were also monitored at 12.2 GHz and show correlated variations. We find the Lomb-Scargle periodogram to be the most sensitive method to search for periodicity but possibly prone to false detections. Periods range from 132.8 days  (with 26 cycles observed) to 509 days (with 7 cycles observed).  Five of the sources show arguably periodic variations, while G331.13-0.24 shows strong periodicity in one peak, with large and variable delays in other peaks.

\end{abstract}


\label{firstpage}

\section{Introduction}

Class II methanol masers are well established as tracers of an early stage of massive star formation \citep[][and references therein]{Ellingsen2006}. The most prevalent and strongest methanol maser line is the $5_1-6_0A^+$ transition at 6.7 GHz, first discovered by \citet{Menten1991}.  To date, over 1000 6.7 GHz methanol masers are known \citep{Pestalozzi2005,Green2009}.  The $2_0-3_{-1}E$ transition at 12.2 GHz is not as common or as intense but is generally closely associated with the 6.7 GHz masers \citep{Breen2011a}. \citet{Breen2011a} speculate that sources that have both 6.7 and 12.2 GHz methanol masers may be slightly more evolved than those with only 6.7 GHz masers. The location of the masers in relation to the newly formed star is still not well understood.  High resolution observations have shown masers to have a variety of morphologies \citep{Norris1993,Phillips1998,Walsh1998,Minier2002,Dodson2004}, including linear and curved structures and even a perfect ring shape \citep{Bartkiewicz2005}.  Some of these structures have been explained as parts of edge-on or inclined disks \citep{Norris1993,Minier2000,Pestalozzi2004}, in outflows \citep{Minier2000} or associated with shocks \citep{Walsh1998,Dodson2004}.

The high intensity of the methanol masers enables them to be easily monitored with smaller radio-telescope dishes, which typically have more time available for long-term monitoring programmes.  Masers are expected to be extremely sensitive to changes in their environment, including local conditions in the volume of masing gas (which may affect the maser path length), as well as the radiation field.  Class II methanol masers are believed to be pumped by infrared radiation from warm dust \citep{Sobolev1997,Cragg2002,Cragg2005} and the masers amplify the continuum emission from background HII regions.  Thus monitoring flux density variations in masers could lead to new insights into massive star formation but interpretation of the variation is complicated by the dependence of the maser intensity on many external factors.

Variability in class II methanol masers was first identified by \citet{Caswell1995}.  \citet{Goedhart2004} conducted an extensive monitoring program for four years finding several sources that exhibited periodic or quasi-periodic variations.  The source G9.62+0.20E was the first confirmed periodic maser \citep{Goedhart2003} exhibiting simultaneous flares at 6.7, 12.2 and 107 GHz \citep{VanderWalt2009}.   The source G12.89+0.49 was found to exhibit rapid variations with a period of 29.5 days. Over 100 cycles were observed in the Hartebeesthoek monitoring program and the data for this source were published separately \citep{Goedhart2009}.   To date, only two other periodically variable sources have been detected: IRAS 18566+0408 which also shows correlated variability in a formaldehyde maser with a period of 237 days \citep{Araya2010} and G22.357+0.066 which has a period of 179 days \citep{Szymczak2011}. 

The range of periods found (between 29.5 to 509 days) is inconsistent with stellar pulsations of main sequence stars, for which periods range between a few hours to 3 days at the most \citep{Moffat2012}. Likewise, stellar rotation rates for massive protostars range from a few days up to 16 days \cite{Nordhagen2006}. Recently, \citet{Inayoshi2013} proposed that periods in the observed range  could be explained by pulsation of massive protostars undergoing rapid mass accretion. However, their model does not explain the asymetric flare profiles seen in several of the masers.  On the other hand, binary systems could also explain the range of periods detected.  A large fraction (69\%) of main sequence binary stars are found in binary systems and this is expected to be a good representation of their properties at birth \citep{Sana2012}.  \citet{Apai2007} surveyed the radial velocities of 16 embedded massive stars and  found that  20\% of their targets were close binaries.  There are a number of ways in which a binary system could modulate maser emission.  For example, \citet{Muzerolle2013} have detected periodic variations in the infrared luminosity of a young protostar due to pulsed accretion, modulated by a binary companion.  On the other hand,  \citet{VanderWalt2011} suggests that a colliding-wind binary (CWB) system can explain both the range of periods and the flare profiles of most of the masers. In the CWB model, the maser flares are due to changes in the background free-free emission caused by a pulse of ionizing radiation passing through partially ionized gas. The pulse of radiation is produced by shocked gas from stellar winds when the stars are at periastron.  This model can explain the raipid rise of maser intensity as well as the slower decay when the hydrogen recombines. 

In this paper we present the results of approximately 10 years of monitoring of six longer-period sources which have been confirmed to be (quasi-)periodic, from the inception of the monitoring programme up to the failure of the main polar shaft bearing in October 2008.

\section{Observations}

All observations were made with the Hartebeesthoek Radio Astronomy Observatory (HartRAO) 26 m telescope.  Observing took place at one to two week intervals, with observations at two to three day intervals if a source was seen to be flaring. The antenna surface at this time consisted of perforated aluminium panels. The surface was upgraded to solid panels during 2003--2004, resulting in increased efficiency at higher frequencies. However, the telescope pointing was affected until the telescope was rebalanced in 2005 April and a new pointing model was derived and implemented in 2005 September. Pointing checks during observations were done by observing offset by half a beam-width at the cardinal points and fitting a two-dimensional Gaussian beam model to the observed peak intensities. The sources were observed with hour angle less than 2.3 h to minimize pointing errors. Pointing corrections after the implementation of the new pointing model were typically around 4 to 10 per cent at 12.2 GHz and 1 to 5 per cent at 6.7 GHz. Any observations with pointing corrections greater than 30 per cent were excluded from the final data set since it was found that these were invariably outliers in the time series.

Amplitude calibration was based on regular drift scan measurements of  3C123, 3C218 and Virgo A (which is bright but partly resolved), using the flux scale of Ott et al. (1994).  Pointing errors in the north-south direction were measured via drift scans at the beam half power points and the on-source amplitude was corrected using the Gaussian beam model.

During 2003 the receivers were upgraded to dual polarization, and observations were switched to a new control system and a new spectrometer. Frequency-switching mode was used for all observations. 

The observing parameters of the monitoring programme are given in Table 1. Prior to 2003 only left circular polarization was recorded and two different bandwidths were used at 6.7 GHz, depending on the target source's velocity range.

The strong methanol maser source G351.42+0.64 was used as a comparison to identify potential periods induced by the telescope systems.  No evidence of periodicity was found. Detailed analysis of this source was presented in \citet{Goedhart2009} and will not be repeated in this paper.

\begin{table}
\begin{center}
\caption{Observing parameters. Average system temperatures and rms noise are given.}
\label{tab:instr}
\begin{tabular}{lrrrrr}
\hline
Observation dates	&	BW	&	chan.	& vel. res 		&	T$_{sys}$	&	rms	\\
			&	(MHz)	&		&	km s$^{-1}$	& (K)		&	(Jy)	\\
\hline
\multicolumn{6}{c}{Rest frequency: 6.668518 GHz}\\
1999/01/17 -- 2003/04/03	&	0.64	&	256	& 0.112	&51	&0.5		\\
                                         &  0.32  & 256  & 0.056 & 51 & 0.5 \\
2003/07/04 -- 2008/09/30	&	1.00	&	1024	&0.044	&70	&0.4		\\
\\
\multicolumn{6}{c}{Rest frequency: 12.178593 GHz}\\
2000/01/30 -- 2003/04/07	&	0.64	&	256	&0.062	&139	&2.0		\\
2003/08/25 -- 2008/08/06	&	1.00	&	1024	&0.048	&99	&0.3		\\
\hline
\end{tabular}
\end{center}
\end{table}

\section{Period search methods}

\begin{figure}
\resizebox{\hsize}{!}{\includegraphics[clip]{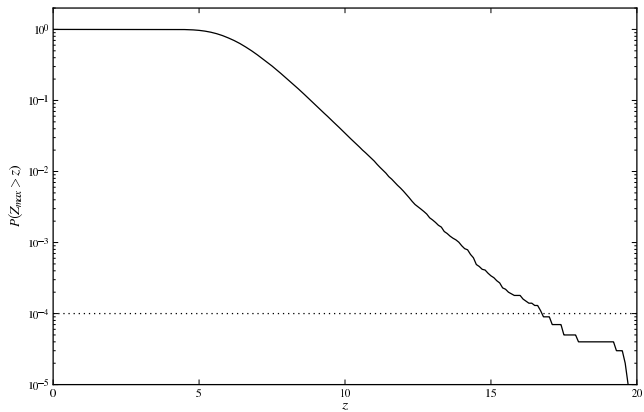}}
\caption{The cumulative density function for the maximum power found for periodograms of synthetic noise for G9.62+0.20 at 6.7 GHz. }
\label{fig:CDF}
\end{figure}

There are many methods of searching for periodicities in unevenly sampled astronomical data, most of which are variations on Fourier transforms  \citep[eg][]{Scargle1989} or folding data by a trial period and measuring the dispersion of the data points through a test statistic \citep[eg][]{Stellingwerf1978}. We have tried a number different methods of detecting periodicity and found the Lomb-Scargle periodogram, using the fast algorithm of Press and Rybicki (1989), to be the most sensitive. However, this method can give rise to false detections.   We follow the method recommended by \citet{Frescura2008} to determine the false alarm probability function. A cumulative density function (CDF) was constructed from 100000 Monte Carlo simulations of time series of  gaussian noise, from which the maximum power in the Lomb-Scargle periodogram was found.  The mean and variance of the noise was estimated from emission-free channels in the spectra and the same timestamps as the observations were used. CDFs were calculated individually for each source and transition. In the cases of  G9.62+0.20 and G328.24-0.55 the two observational phases had different noise distributions. This was reflected in the synthetic time-series. The effect of oversampling was also investigated, using a conservative factor of 4, and grossly oversampling at a factor of 20.  The CDFs are very similar in all cases. Frequencies with false alarm probability $\leq$1e-4 were considered to be significant. Figure~\ref{fig:CDF} shows the CDF for G9.62+0.20 at 6.7 GHz.  We conclude from this that a power level of 16.7 corresponds to a false-alarm probability of 0.01\%.

Data were typically detrended using an unconstrained first or second-order polynomial prior to period search.  Despite this, low-frequency compononents were still found in the data, particularly for some of the maser peaks in G9.62+0.20. \citet{Kidger1992} discuss the problem of confirming periodicity with the intent to predict future behaviour.  They state that the data sample should be of very long duration, covering at least six cycles, and the variations should be of large amplitude.  More cycles are necessary to confirm periodicity in the case of objects with low amplitude.  This is consistent with our own observations.  \citet{Goedhart2004} identified a number of sources which were potentially periodic, but had observed only three  cycles for some objects.  Continued monitoring of G196.45-1.68 (which had shown two and a half sinusoidal cycles) and G316.64-0.09 (three regularly-spaced flares of decreasing amplitude) showed no further evidence of periodic variations.

Epoch-folding was used to verify the frequencies found from the Lomb-Scargle periodogram. It was found that the standard epoch-folding method of finding the maximum $\chi^2$ was not very sensitive to flares of varying amplitude.  The Davies L-statistic \citep{Davies1990} appears to be the most sensitive for the sort of time-series that we have observed. In the case of sources which may not be strictly periodic, it is useful to have an estimate of the uncertainty in the periods.  We investigated methods recommended by \citet{Leahy1987} and \citet{Larsson1996} but these are dependent on being able to accurately model the wave-form of the pulse. Some sources show significant variation in the pulse shape from one cycle to another.  All of the  sources appear to have varying amplitudes in each cycle, which adds an additional free parameter. The width of the peak in the epoch-folded periodogram appears to be the most robust way to reflect the uncertainty in the periods. In the case of quasi-periodic sources, this is not due to measurement errors, but is a reflection of the spread in times at which the flares peak.

\section{Results}

\textbf{\textit{The Lomb-Scargle periodograms will be available online as supplementary material.}}
\subsection{G9.62+0.19E}

G9.62+0.20E is the brightest 6.7 GHz methanol maser known and was observed to reach a peak flux density of 7344 Jy on 24 June 2006.  It has been found to be associated with the HII region E in the massive star forming complex G9.62+0.20 \cite{Garay1993} and was classified as a hypercompact HII region by \citet{Kurtz2002}. The distance to this region from trigonometrical parallax of the 12.2 GHz masers is 5.2 kpc \citep{Sanna2009}. Infall motions with HCN (4--3) and CS (7--6) lines were found towards the submilliter core in region E with an infall rate of $4.3\times10^{-3} M_\odot$ yr$^{-1}$ \cite{Liu2011}. 

\begin{figure}
\resizebox{\hsize}{!}{\includegraphics[clip]{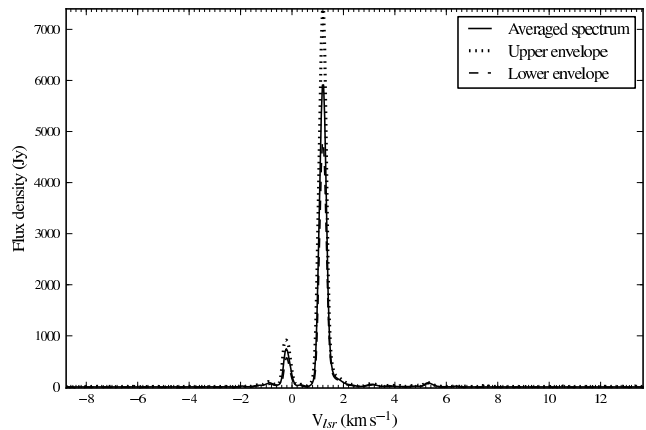}}
\caption{Range of variation across all spectral channels for G9.62+0.19E at 6.7 GHz during 2003--2008.}
\label{fig:g0096_67_spectra}
\end{figure}

\begin{figure}
\resizebox{\hsize}{!}{\includegraphics[clip]{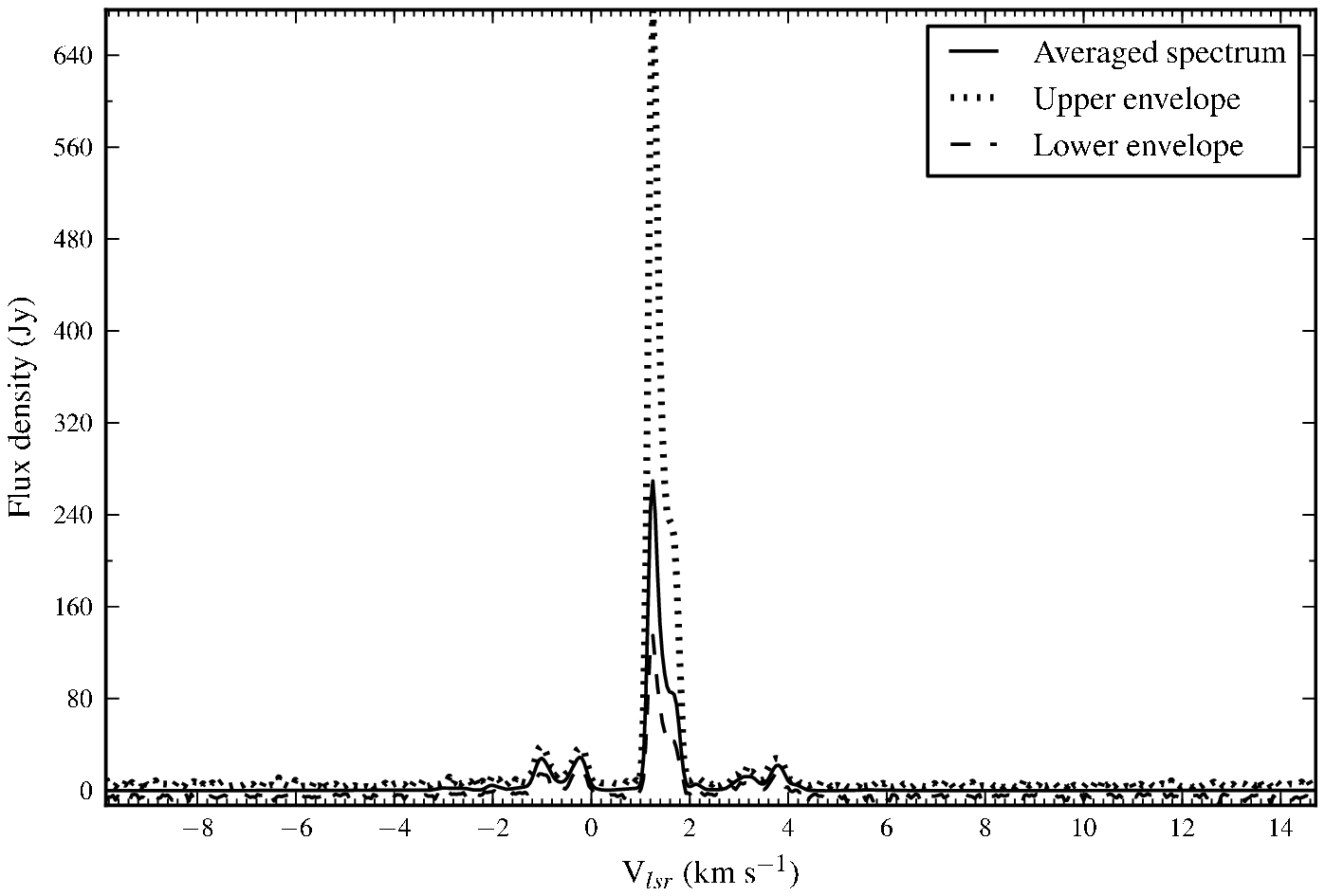}}
\caption{Range of variation across all spectral channels for G9.62+0.19E at 12.2 GHz during 2003--2008.}
\label{fig:g0096_122_spectra}
\end{figure}

\begin{figure}
\resizebox{\hsize}{!}{\includegraphics[clip]{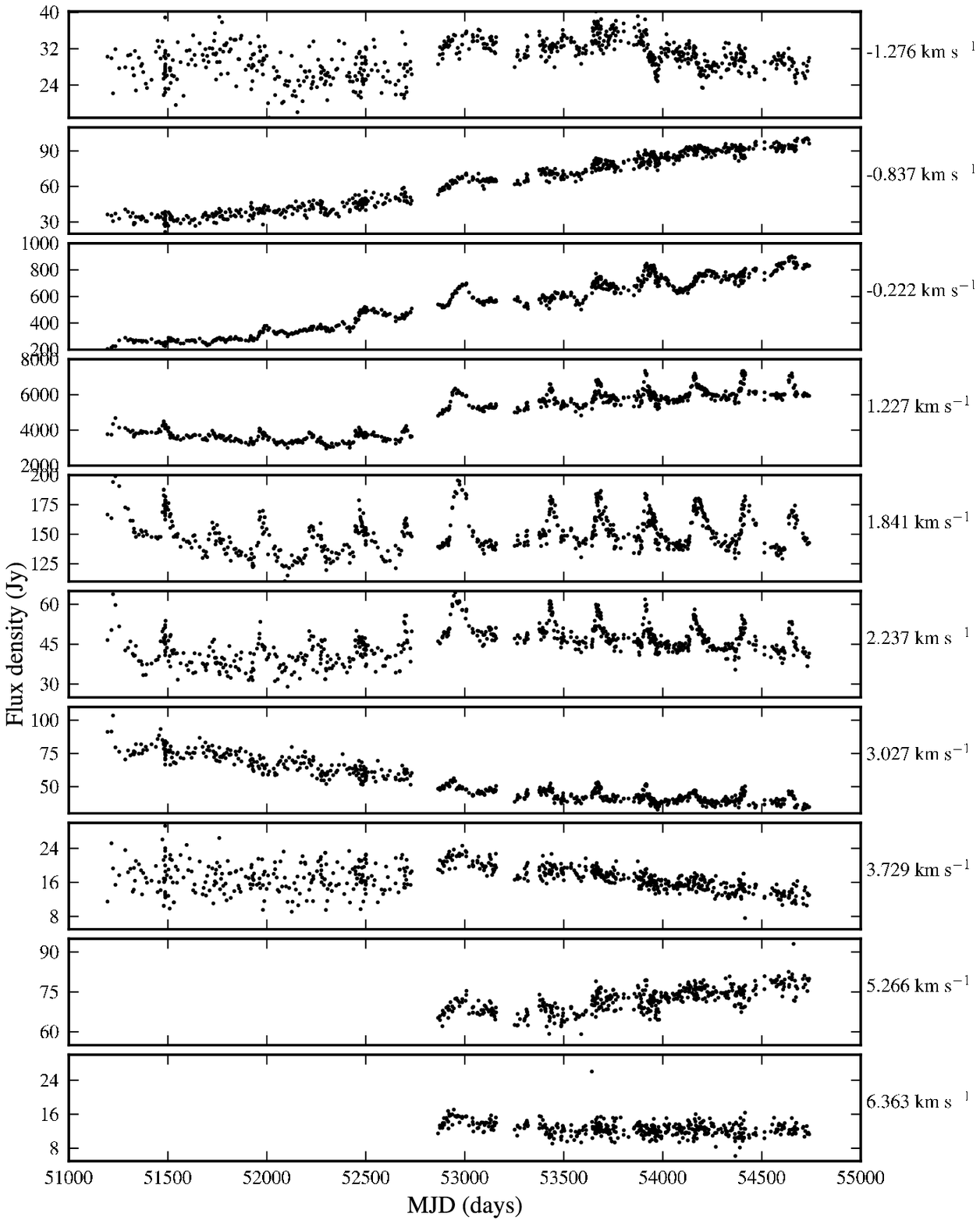}}
\caption{6.7 GHz time series for peak velocity channels in G9.62+0.19E.}
\label{fig:g0096_67_ts}
\end{figure}

\begin{figure}
\resizebox{\hsize}{!}{\includegraphics[clip]{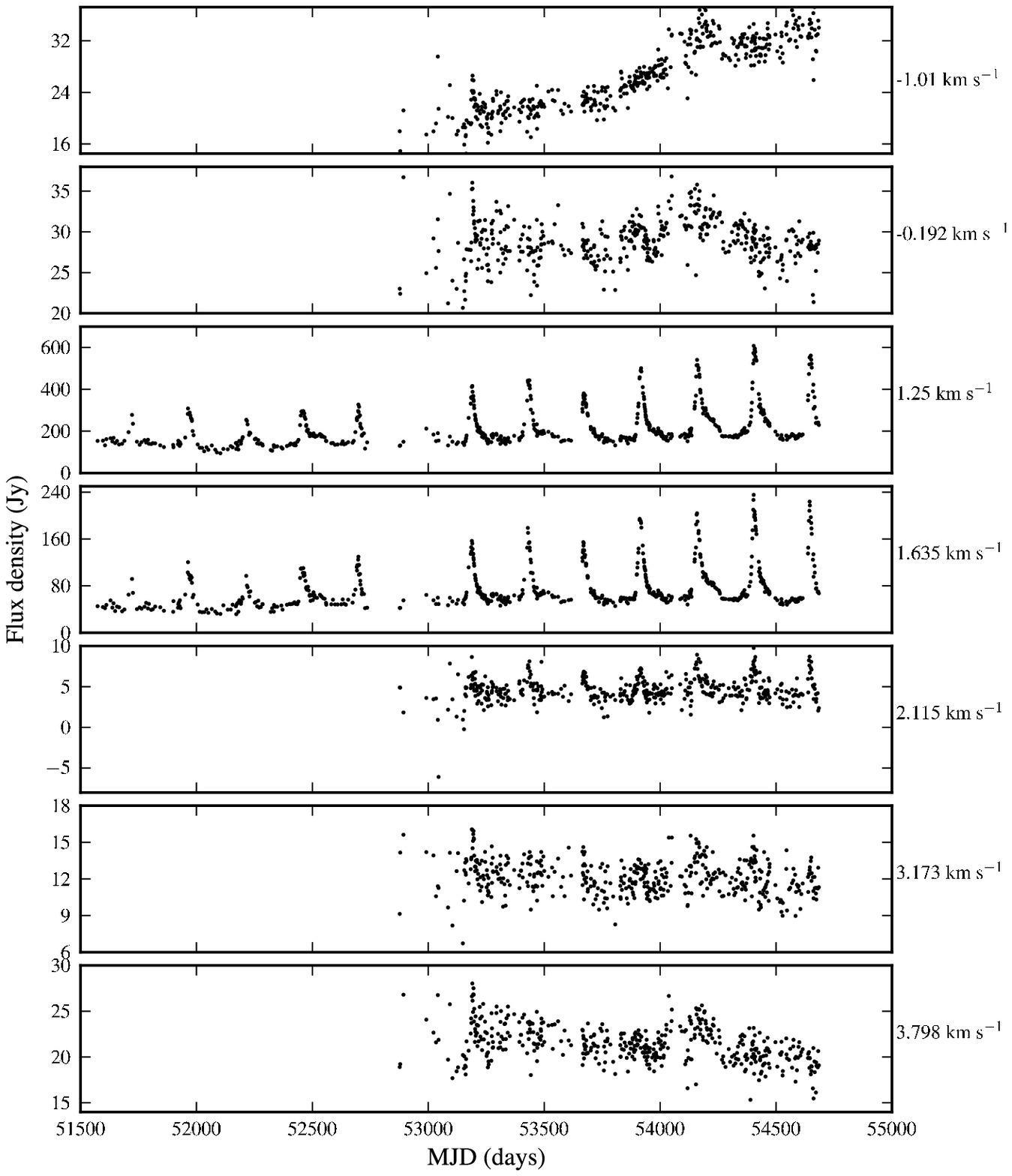}}
\caption{12.2 GHz time series for peak velocity channels in G9.62+0.19E.}
\label{fig:g0096_122_ts}
\end{figure}

The range of variation for all spectral channels for G9.62+0.19E at 6.7- and 12.2 GHz is shown in Figures~\ref{fig:g0096_67_spectra} and \ref{fig:g0096_122_spectra}.  We calculate the maximum, mean and minimum over time for each channel. The time series for the peak channels at 6.7 GHz is shown in Figure~\ref{fig:g0096_67_ts} and at 12.2  GHz in Figure~\ref{fig:g0096_122_ts}. At 6.7 GHz, the components at -0.837, -0.222, 1.227 and 5.266 \kms have been steadily increasing in mean intensity, while the components at 3.027 and 3.729 \kms have been decreasing. The mean brightness of the components at 12.2 GHz have also been increasing slowly but most notable is the increase in the amplitude of the flares in the 1.250 and 1.635 \kms components.

Table~\ref{tab:g962-freqs} summarises the periods found from the Lomb-Scargle periodogram and epoch-folding.  The time-series were detrended using a third-order polynomial.   Several low-frequency components were found with high power, corresponding to long-term trends in the data. Following the criteria of \citet{Kidger1992} we consider a period to be confirmed only if at least six cycles have been observed.   The period with the highest power, for any of the spectral features at 6.7 GHz, is 244.4 days. Clear periodicity is seen in the features at 1.227, 1.841, 2.237 and 3.027 \kms. A harmonic series is seen in the periodograms for 1.841 and  2.237 \kms. At 12.2 GHz the period with the highest power is 244.0 days. Clear periodicity is seen in the same velocity range as for the 6.7 GHz transition and several features show a harmonic series.  The large number of significant frequencies found seem unlikely to be real detections. The periods were investigated further using epoch-folding. Clear peaks are only seen for periods corresponding to $\sim$ 243 days and multiples thereof.   Similar results are seen for the 12.2 GHz data.  Taking the weighted mean of the epoch-folded periods we find a period of 243.3$\pm$2.1 days. The time-series were folded modulo the other periods derived from the Lomb-Scargle periodogram but no clear waveform was seen.

\begin{figure}
\resizebox{\hsize}{!}{\includegraphics[clip]{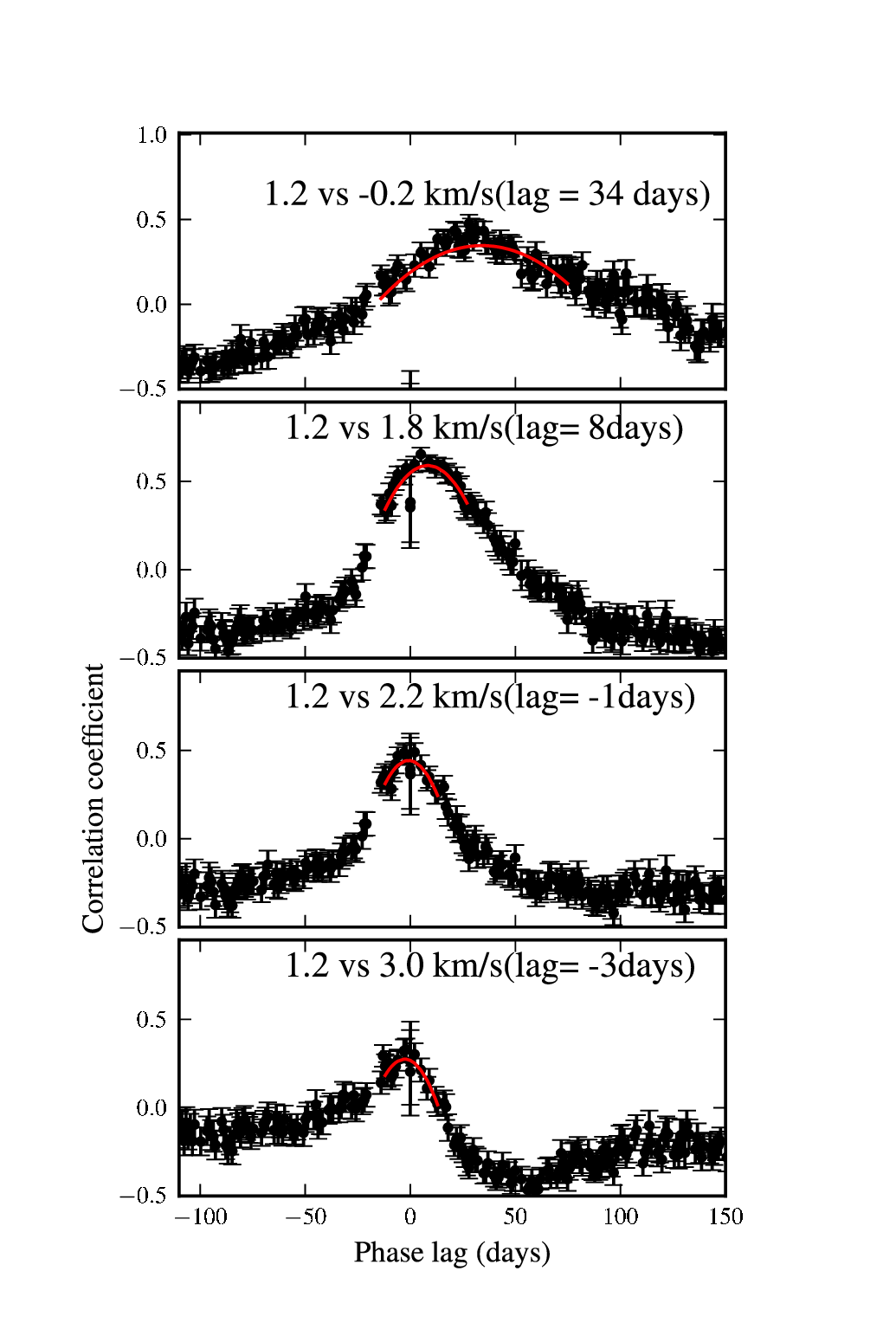}}
\caption{Discrete correlation function between pairs of features at 6.7 GHz for G9.62+0.19E.} 
\label{fig:g0096_67_dcf}
\end{figure}

Delays in flaring between features were estimated using the z-transformed discrete correlation function of \citet{Alexander1997}. We use the 1.227 \kms feature as the reference.  At 6.7 GHz there appears to be a lag of 34 days between the the dominant peak and the -0.222 \kms feature (Figure~\ref{fig:g0096_67_dcf}). There may be an 8 day lag between 1.183 and 1.841 \kms. There are smaller lags between the main peak and 2.2 and 3.0 \kms but these may not be significant given the noise in the correlation function. No lags are found between the features in the 12.2 GHz time series and there is no lag between the same velocity features at 12.2 and 6.7 GHz.

\begin{table*}
\caption{Periods found from Lomb-Scargle periodogram and epoch-folding for G9.62+0.19E. Identified harmonic series are indicated in bold text.}
\label{tab:g962-freqs}
\begin{center}
\begin{tabular}{lllllll}
\hline
Velocity & mean & mean & S/N & L-S & E-F period & E-F  \\
                & flux 	&		rms	&		&	Significant periods & & HWHM   	\\
                &	density	&	noise		&		&		& & 	\\
\kms     & 	Jy							& Jy				&					& Days      & Days & Days                    \\
\hline
\multicolumn{7}{c}{6.7 GHz}\\
-1.276   & 30.1						& 1.9			&  15			& 1575, 945.2, 683.2, 187.8 & -- & -- \\
-0.837   & 65.7						& 2.2			& 30				& 1013 & -- & --\\
-0.222   & 564.5					& 8.4			& 67				& 821.9, 443.0, 241.3, 142.1, 131.0 & -- & --\\
1.227    & 5079						& 75				&	637			& 1620, 756.1, 446.5, 268.8, \textbf{244.4,127.4, 81.0, 61.0} & 243.8 & 4.1\\
1.841    & 151.6					& 3.4			& 45				& 450.1, 272.6, \textbf{244.4}, 167.8, 128.0, \textbf{121.7, 81.2, 61.0 } & 243.6 & 3.8\\
2.237    & 45.6						& 2.1			& 22				& 1890, \textbf{243.4}, 128.0, \textbf{122.0, 81.2, 61.0} & 243.7 & 3.4 \\
3.027    &52.0						& 2.5			& 23				& \textbf{246.6}, 225.0, \textbf{122.2, 81.7} & 242.6 & 5.5 \\
3.729	& 17.0						& 1.9			& 9				& 1829	& -- & --	\\
5.266    & 72.0						& 1.4			&	51				& 394.7 & -- & --\\
6.363    & 12.6						& 1.1			& 11				&  -- & -- & --\\
\multicolumn{7}{c}{12.2 GHz}\\
-1.010	& 26.6						& 0.9			& 30				& 	1071		& -- & --	\\ 
-0.192	& 28.9						&	0.9			&	32				& 997.5		& -- & --\\
1.250	& 277.5						& 1.8			& 124			&  \textbf{244.0, 121.7, 81.1, 60.9, 48.7, 40.6} & 243.2 & 3.6\\
1.635	& 79.2						& 1.3			& 60				&  \textbf{244.0, 121.7, 81.1, 60.9, 48.7, 40.6} & 243.5 & 2.2\\
2.115	& 4.7						& 0.9			& 5				&  \textbf{245.2, 120.0, 81.3, 60.6} & 243.3 & 6.5\\
3.173	& 12.18						& 0.9			& 14				&  \textbf{80.4} & 241.6 & 5.5\\
3.798	&	21.5						&	0.9			&	24				& 803.4, 140.4 & -- & -- \\
\hline
\end{tabular}
\end{center}
\end{table*}

\subsection{G188.95+0.89}

These masers are situated in the star forming region AFGL 5180 or S252.  VLBI observations at 6.7 and 12.2 GHz show a linear distribution, with the two transitions spatially co-located \citep{Minier2000}. No continuum radio source has been detected, but the masers are projected on a bright mm source with an estimated mass of 50 M$_\odot$  \citep[MM1 in][]{Minier2005}. \citet{Longmore2006}  estimate the mass of the mid-infrared core associated with the methanol masers to be 7 M$_\odot$ based on its luminosity of 8.4 L$_\odot$. Parallax measurements of water masers associated with the star forming region place it at a distance of 2.02 kpc \citep{Niinuma2011}.

\begin{figure}
\resizebox{\hsize}{!}{\includegraphics[clip]{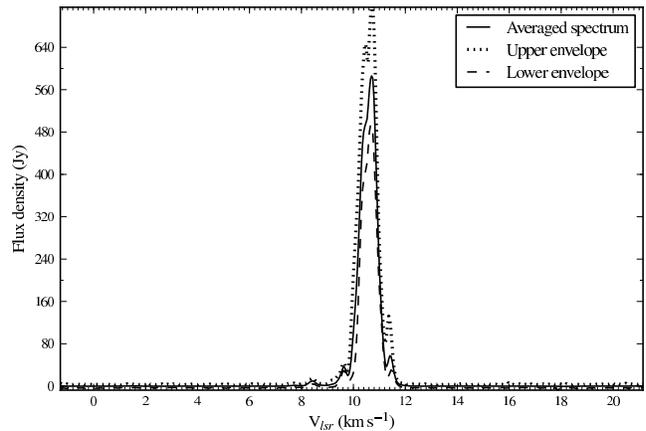}}
\caption{Range of variation across all spectral channels for G188.95 at 6.7 GHz during 2003--2008.}
\label{fig:g1889_67_spectra}
\end{figure}

\begin{figure}
\resizebox{\hsize}{!}{\includegraphics[clip]{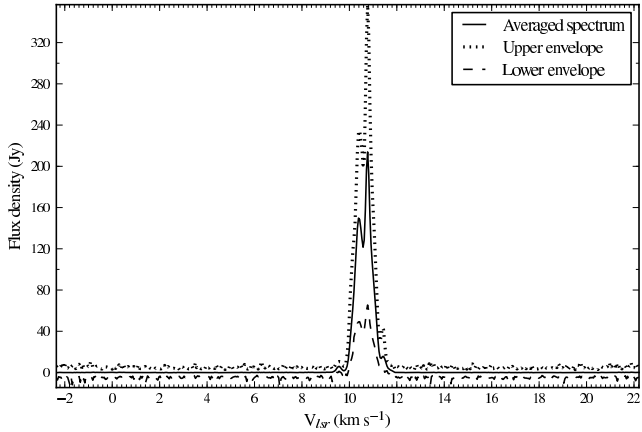}}
\caption{Range of variation across all spectral channels for G188.95+0.89 at 12.2 GHz during 2003--2008.}
\label{fig:g1889_122_spectra}
\end{figure}

\begin{figure}
\resizebox{\hsize}{!}{\includegraphics[clip]{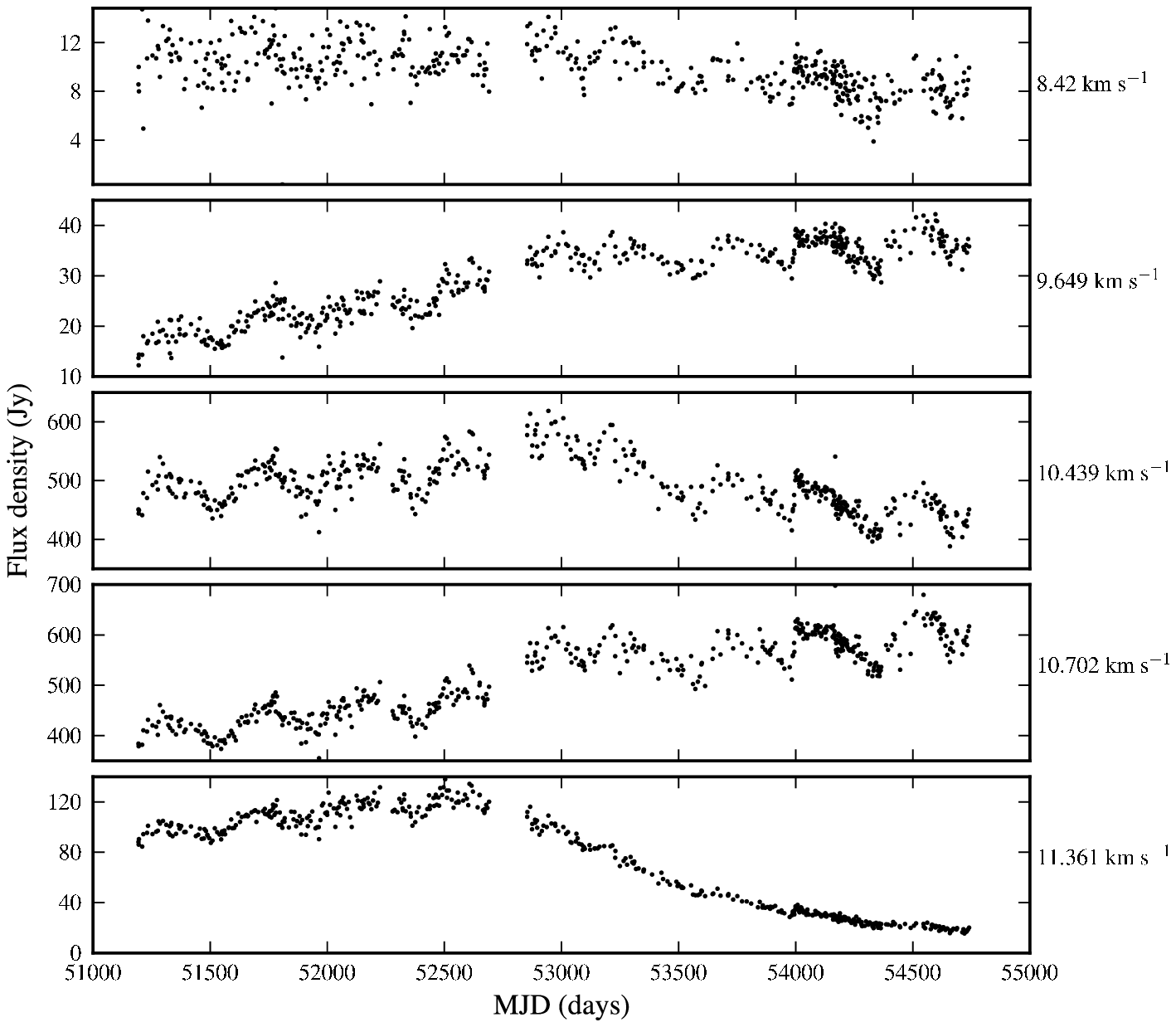}}
\caption{6.7 GHz time series for peak velocity channels in G188.95+0.89.}
\label{fig:g1889_67_ts}
\end{figure}

\begin{figure}
\resizebox{\hsize}{!}{\includegraphics[clip]{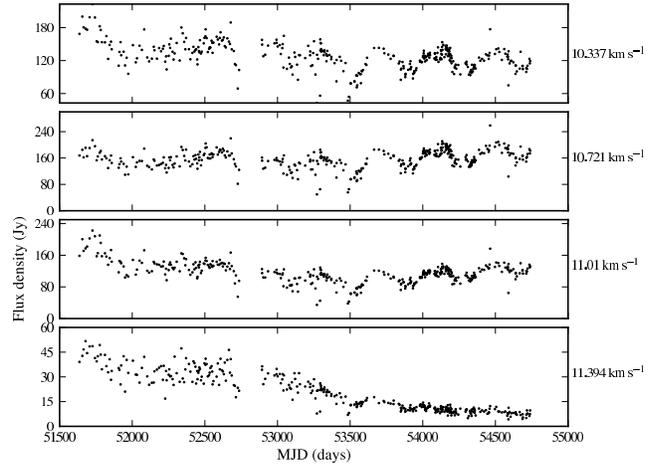}}
\caption{12.2 GHz time series for peak velocity channels in G188.95+0.89.}
\label{fig:g1889_122_ts}
\end{figure}

Figures~\ref{fig:g1889_67_spectra} and \ref{fig:g1889_122_spectra} show the range of variation in the spectra at 6.7 and 12.2 GHz, respectively.  The time series are shown in Figures~\ref{fig:g1889_67_ts} and \ref{fig:g1889_122_ts}. As with G9.62+0.20, the variations are much more pronounced at 12.2 GHz. The first five cycles at 6.7 GHz show a sinusoidal waveform, but the last four cycles show more of a sawtooth pattern. The feature at 11.361  \kms has been steadily weakening since early 2004 but weak periodic variations can be  seen. \cite{VanderWalt2011} explains this decay by recombination of the ionized gas in part of the HII region along the line of sight to this maser.

Two separate polynomials were used to detrend the time series for the feature at 11.3 \kms prior to period search. The periods above the significance threshold for the Lomb-Scargle periodogram and the periods found from epoch-folding are summarised in Table~\ref{tab:g1889-freqs}. The epoch-fold peaks at 12.2 GHz were asymmetrical and noisy, so they have not been fitted. Using the 6.7 GHz values only we find a weighted mean period of 395$\pm$8 days.  No phase lags between features were found.

\begin{table*}
\caption{Periods from Lomb-Scargle periodogram and epoch-folding  for G188.95+0.89.}
\label{tab:g1889-freqs}
\begin{center}
\begin{tabular}{lllllll}
\hline
Velocity & mean & mean & S/N & L-S & E-F period & E-F  \\
                & flux 	&	rms		&		&	Significant periods & & HWHM   	\\
                &	density	&	noise		&		&		& & 	\\
\kms     & 	Jy							& Jy				&					& Days      & Days & Days                    \\
\hline
\multicolumn{7}{c}{6.7 GHz}\\
8.420	& 9.8				& 1.1		& 9			& 399.6  & 393.9 & 31.0 \\
9.649	& 29.7				& 1.7		& 25		&  1669, 468.9, 394.0 & 396.0 & 16.4 \\
10.439	& 490.0				& 9.6		& 51		& 1830, 886.4, 472.8, 396.8 & 393.9 & 15.8 \\
10.702	& 520.0				& 9.5		& 55		&  1773.1, 915.2, 472.8, 396.8 & 396.3 & 14.5\\
11.361	& 71.1				& 2.1		& 33		&  1830, 411.2 & 399.6, 395.9, 21.8 \\
\multicolumn{7}{c}{12.2 GHz}\\
10.337	& 127.4				& 1.7		& 76		&   1418, 394 &  -- & --\\
10.721	&	160.0			& 1.8		& 89		&  1418, 813, 391 & -- & -- \\
11.010  &	157.7			&  1.6		& 70		& 1418, 827, 387 & --&--\\
11.394	& 	19.6			&	1.4		& 14		& 	1504 & -- &--\\
\hline
\end{tabular}
\end{center}
\end{table*}

\subsection{G328.24-0.55}

There are two maser groups in the beam, viz. G328.236-0.547 and G328.254-0.532 \citep{Phillips1998}. G328.236-0.547  is broken up into two groups with velocity ranges -46 to -42 \kms and -37 to -31 \kms.  G328.254-0.547 has features at -50 \kms and at -40 to -36 \kms.  The latter group is unfortunately blended with G328.254-0.532 and the spectra are probably dominated by the emission from this group. The two velocity groups in G328.236-0.547 are separated by 83 arcsec and lie on either side of an unresolved HII region.  \citet{Phillips1998} offer three explanations for this morphology: shock fronts in a bipolar outflow, two clusters on either side of a thick disk, or two separate sources in a binary system. \citet{Dodson2004} imaged this region with the LBA and find that the maser  spots may have a linear distribution. However, they find no reason to think that the two regions could be related since there is no alignment in position angle of the maser spots. These regions are estimated to be at the near kinematic distance of 3 kpc.

\begin{figure}
\resizebox{\hsize}{!}{\includegraphics[clip]{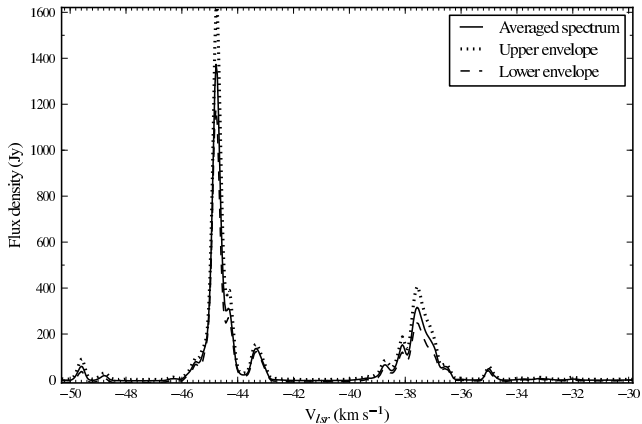}}
\caption{Range of variation across all spectral channels for G328.24-0.55 at 6.7 GHz during 2003--2008.}
\label{fig:g3282_67_spectra}
\end{figure}

\begin{figure}
\resizebox{\hsize}{!}{\includegraphics[clip]{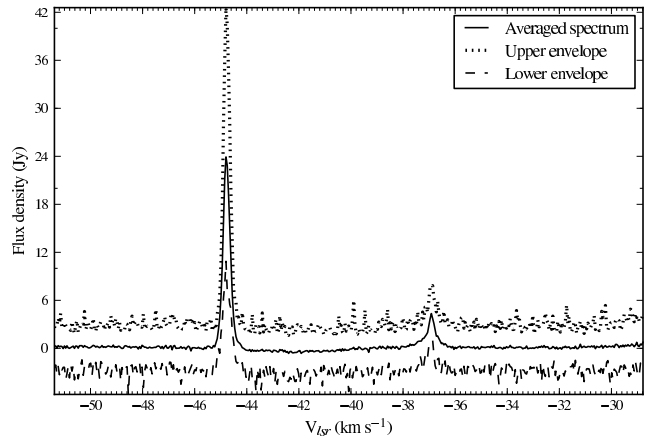}}
\caption{Range of variation across all spectral channels for G328.24-0.55 at 12.2 GHz.}
\label{fig:g3282_122_spectra}
\end{figure}

Figures~\ref{fig:g3282_67_spectra} and \ref{fig:g3282_122_spectra} show the range of variation in the spectra at 6.7 and 12.2 GHz respectively.  There is only one spectral feature at 12.2 GHz strong enough for analysis.   We see correlated periodic variations across several peaks corresponding to the velocity range covered exclusively by G328.236-0.547. The overall intensity of the peak at -44.751 \kms has increased almost threefold since the start of the monitoring programme. The average intensity of the features corresponding to G328.254-0.532 do not show the same trend. The 6.7 GHz periodograms are dominated by long-term trends.   Epochfolding after a second-order detrend finds a  weighted mean period of 220.5$\pm$1.0 days for five features -- -45.410, -44.751, -44.268, -43.259 and -36.410 \kms. The significant periods found in the Lomb-Scargle periodogram and best-fit periods from epoch-folding are summarised in Table ~\ref{tab:g3282-freqs}. The highest spectral power is found for a period of 220.3 days.  The 12.2 GHz epoch folded periodogram is not as sharply delineated as at 6.7 GHz because of the shorter time series.  We adopt a  weighted mean period of 220.5$\pm$1.0 days.  The changing amplitudes of the flares leave the peaks weakly defined. The features corresponding to G328.254-0.547 show correlated, regular variations with a characteristic period close to 300 days but the epoch-fold does not show peaks corresponding to those found by the Lomb-Scargle periodogram and no clear waveform is seen when the time series are folded modulo 300 days. The variations show no correlation with G328.236-0.547. No time-delays were found.

\begin{table*}
\caption{Periods  from Lomb-Scargle periodogram and epoch-folding for G328.24-0.55}
\label{tab:g3282-freqs}
\begin{center}
\begin{tabular}{lllllll}
\hline
Velocity & mean & mean & S/N & L-S & E-F period & E-F  \\
                & flux 	&	rms		&		&	Significant periods & & HWHM   	\\
                &	density	&	noise		&		&		& & 	\\
\kms     & 	Jy							& Jy				&					& Days      & Days & Days                    \\
\hline
\multicolumn{7}{c}{6.7 GHz}\\
-49.581 	& 41.3 						& 2.1     		&  20 			& 1548, 311 & -- & --\\
-45.410	&55.8						& 2.2			&	25				& 1639, 978, 220.3 & 220.0 & 5.2\\
-44.751	& 805.4						& 12.4			&	65				&  1689, 221.1 & 221.5 & 3.5 \\
-44.268	& 222.0						& 4.0			&  55			& 220.3, 189.6 & 220.0, 5.7\\
-43.259	& 106.8						& 2.7			& 40				& 1798, 220.3 & 219.9 & 5.3 \\
-38.649	& 56.1						& 2.2			& 25				& 1639, 395, 309.7 & -- & --\\
-37.990	& 145.4						& 3.3			& 44				& 1548.3, 392.5, 352.8, 294.9 & -- & --\\
-37.420	& 308						& 5.7			& 55				& 1506, 389.8, 302.9 & -- & --\\
-37.068	& 185.9						& 4.0			&	47				& 1429, 290.3 & -- & --\\
-36.410	& 43.8						& 2.1			& 21				& 1394, 381.8, 218.6 & 220.8 & 8.4\\
\multicolumn{7}{c}{12.2 GHz}\\
-44.798 	& 21.2						& 1.1     		&  20 			& 477.6, 216.3 & -- & --\\
\hline
\end{tabular}
\end{center}
\end{table*}

\begin{figure}
\resizebox{\hsize}{!}{\includegraphics[clip]{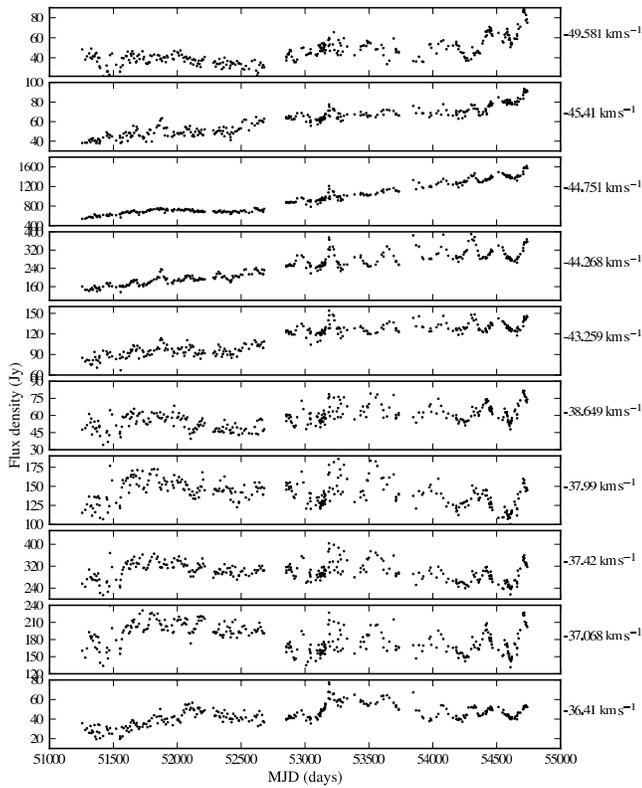}}
\caption{6.7 GHz time series for peak velocity channels in G328.24-0.55.}
\label{fig:g3282_67_ts}
\end{figure}

\begin{figure}
\resizebox{\hsize}{!}{\includegraphics[clip]{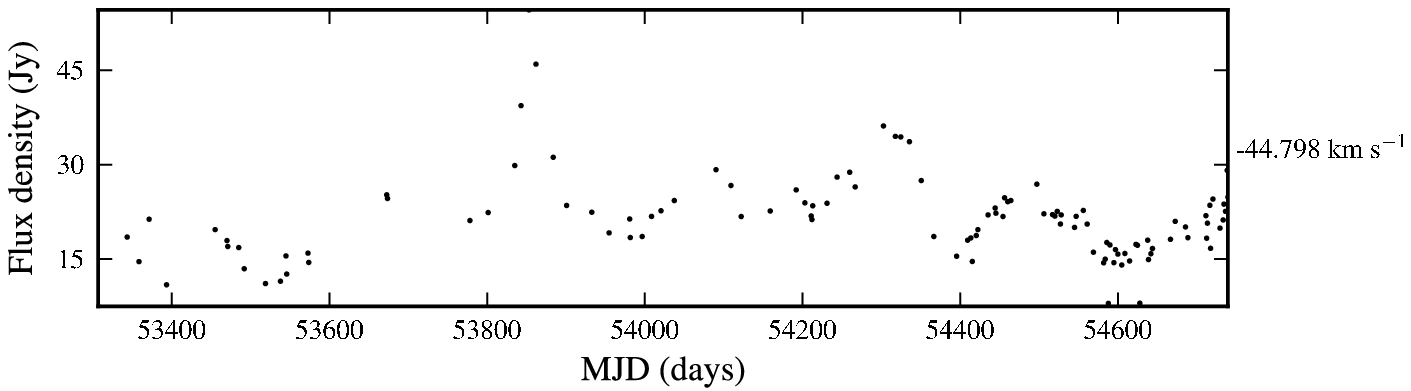}}
\caption{12.2 GHz time series for the peak velocity channel in G328.24-0.55.}
\label{fig:g3282_122_ts}
\end{figure}

\subsection{G331.13-0.24}

The maser spots in this source have a linear distribution with a velocity gradient, except for a single component \citep{Phillips1998}. The masers lie at the edge of an extended HII region with an irregular morphology. \citet{Phillips1998} speculate that there may be more than one star embedded in the UC HII region because the masers are offset from the centre. \citet{DeBuizer2009} included this region in their sample to search for outflows associated with linear maser structures. Using ATCA, they find SiO emission centred on the maser location and distributed at a very similar angle to the masers although the outflow may be oriented along the line of sight. There is a 3mm continuum source which overlaps with the cm continuum source but is slightly offset. 

\begin{figure}
\resizebox{\hsize}{!}{\includegraphics[clip]{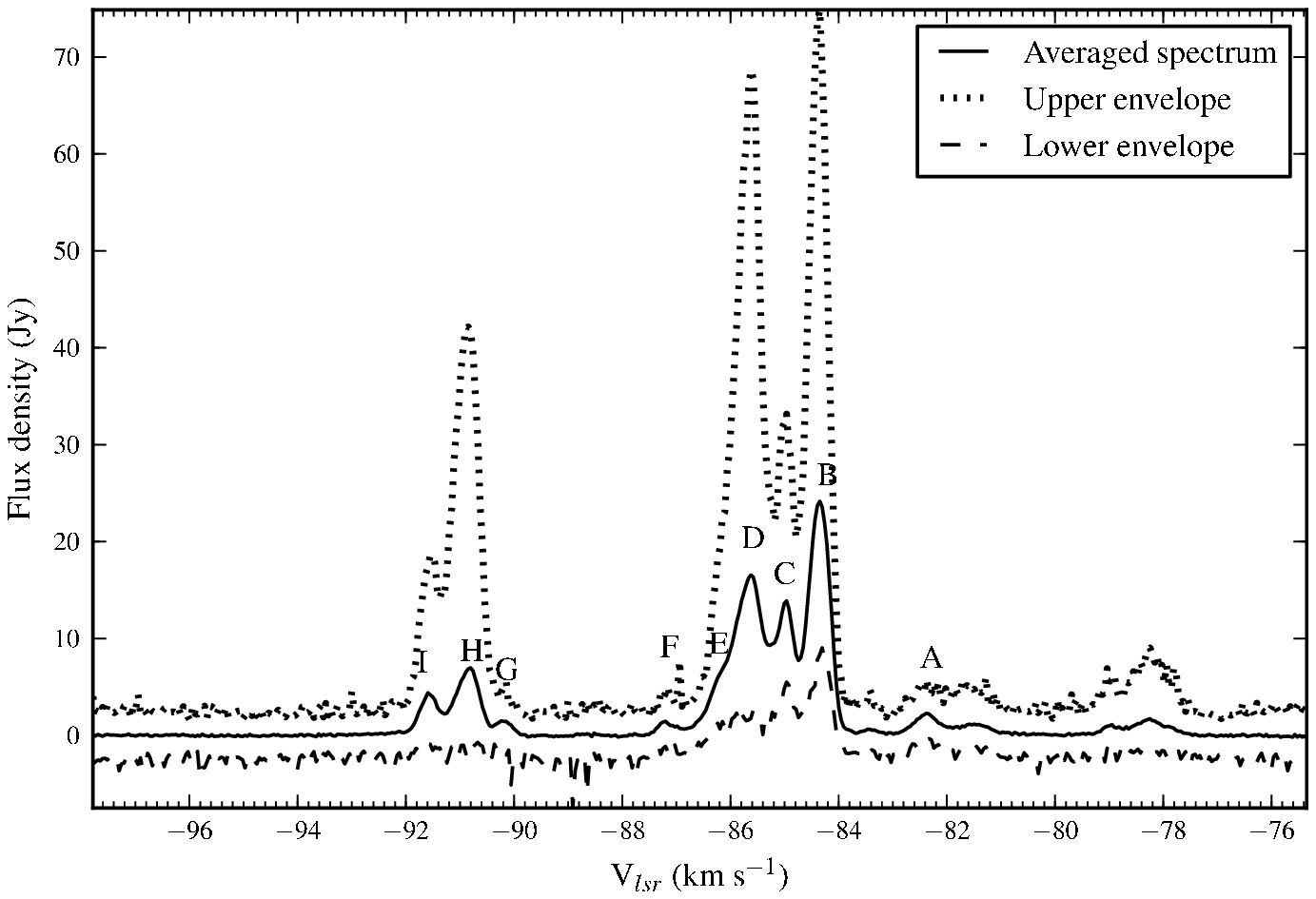}}
\caption{Range of variation across all spectral channels for G331.13-0.24 at 6.7 GHz during 2003--2008.}
\label{fig:g3311_67_spectra}
\end{figure}

\begin{figure}
\resizebox{\hsize}{!}{\includegraphics[clip]{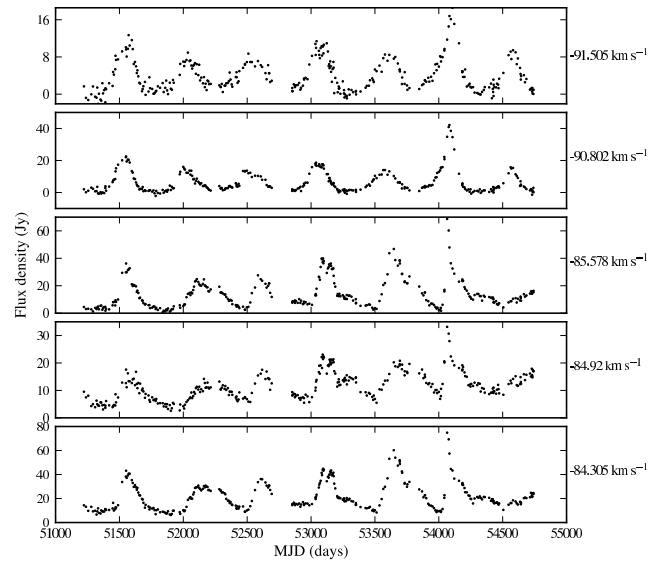}}
\caption{6.7 GHz time series for peak velocity channels in G331.13-0.24.}
\label{fig:g3311_67_ts}
\end{figure}

\begin{figure}
\resizebox{\hsize}{!}{\includegraphics[clip]{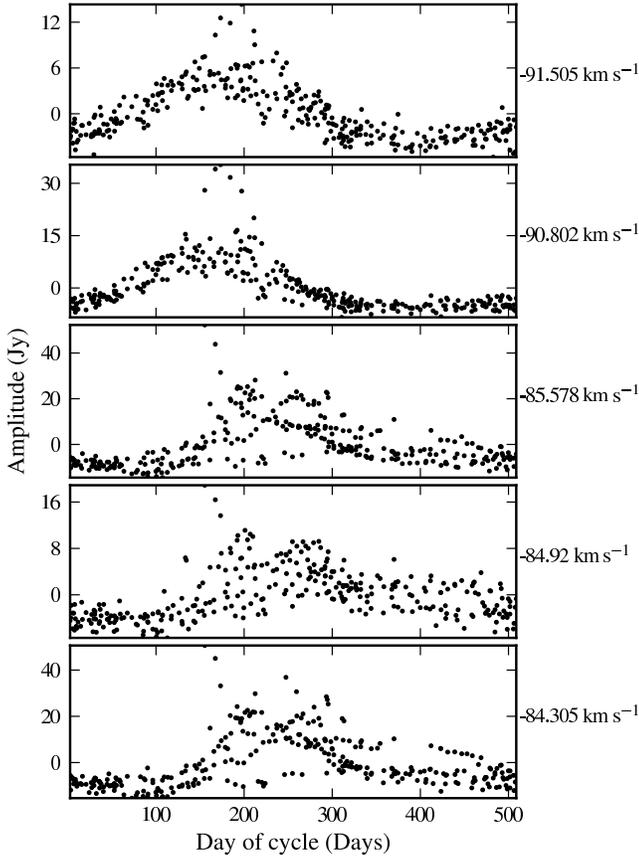}}
\caption{6.7 GHz time series for G331.13-0.24 folded modulo 510 days.  The data have been detrended with a first-order polynomial.} 
\label{g3311_67_fold}
\end{figure}

\begin{figure}
\resizebox{\hsize}{!}{\includegraphics[clip]{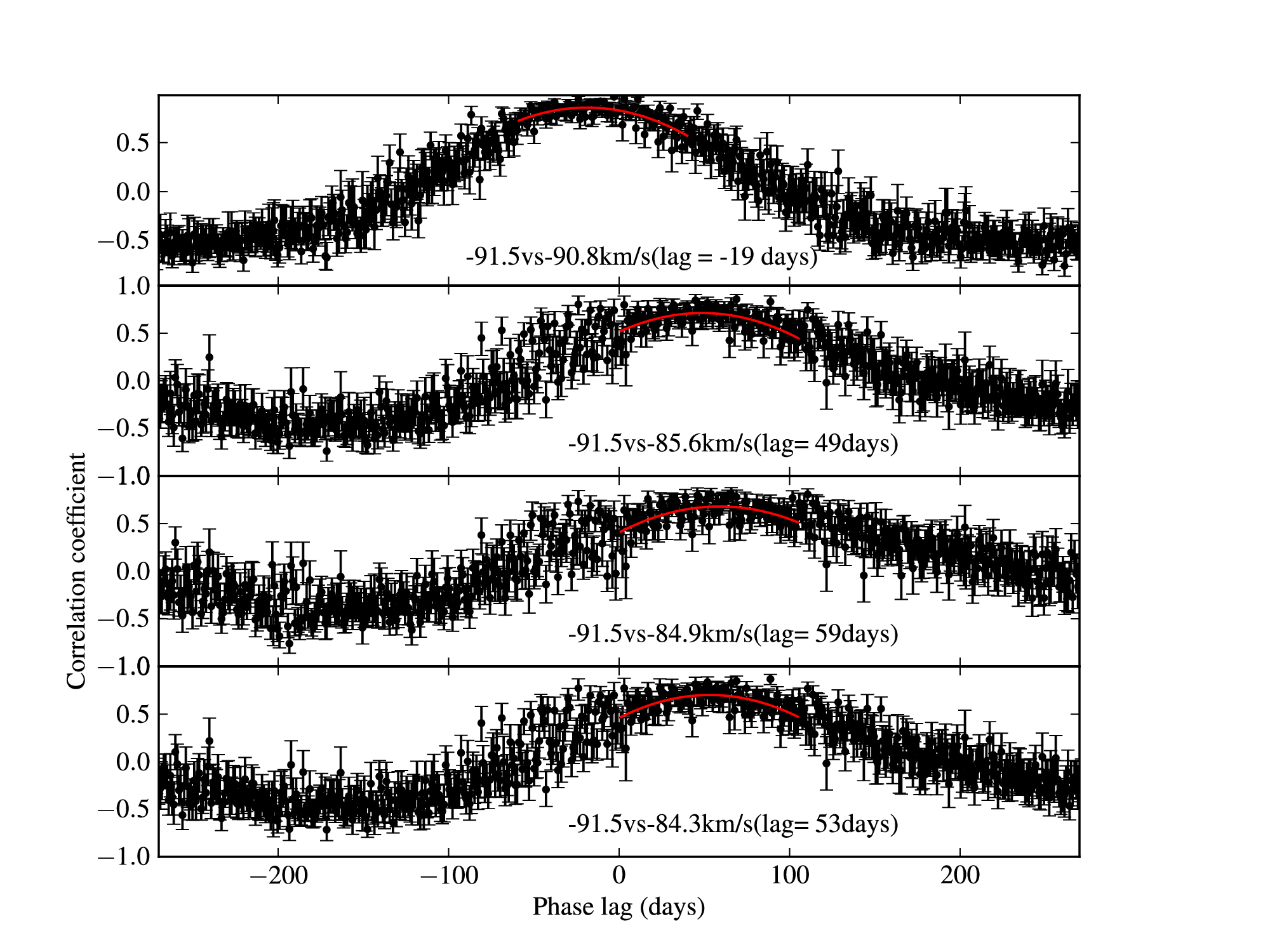}}
\caption{Discrete correlation function between pairs of features for G331.13-0.24.} 
\label{fig:g3311_67_dcf}
\end{figure}

Figure~\ref{fig:g3311_67_spectra} shows the spectra at 6.7 GHz with the spots mapped by \citet{Phillips1998} indicated.  While there are 12.2 GHz masers associated with this source, they lie below the detection threshold of the 26-m telescope.  The 6.7 GHz  time-series are plotted in Figure~\ref{fig:g3311_67_ts}.  All peaks show repeated flares with a delay between the two velocity groupings.  The last flare failed to manifest at the expected time in the second grouping.  The second grouping is also showing a gradual increase in the base level intensity. The Lomb-Scargle periodograms show a spread in periods of about 7 days and Table~\ref{tab:g3311_67-freqs}).  The spread in derived periods may be due to the flares not repeating very well. The epoch-folded peaks are correspondingly very broad. We adopt a weighted mean period of 509$\pm$10 days from the epoch-folded periodograms. In Figure~\ref{g3311_67_fold} we show the time-series folded modulo 509 days. The delays between the peak of the flares can be clearly seen. It is also clear that the the second group of masers (B, C and D) do not flare as regularly as the first group (H and I). Figure~\ref{fig:g3311_67_dcf} shows the discrete correlation function between the reference feature at  -91.505 \kms (spot I) and the other features. The magnitude of the delay is estimated by fitting a second-order polynomial to the peak of the correlation function but the peaks are quite broad and the derived delays should be treated with care.  The changing pulse shapes make it very difficult to identify the exact magnitude of the delay. The feature at H starts to flare at the same time as I, but it appears to peak about 19 days earlier. Feature D flares about 49 days after I, and features C and B flare 59 and 53 days later. Broadly outlined, this implies that the flare propagates from the south-west (closer to the peak of the HII region) to the north-east. The resolution of the SiO maps from ATCA is unfortunately not good enough to determine if the flaring is related to the outflow. We would also need to be able to know the absolute positions of the masers relative to the outflow.

\begin{table*}
\caption{Periods from Lomb-Scargle periodogram and epoch-folding for G331.13-0.24 at 6.7 GHz}
\label{tab:g3311_67-freqs}
\begin{center}
\begin{tabular}{lllllll}
\hline
Velocity & mean & mean & S/N & L-S & E-F period & E-F  \\
                & flux 	&	rms		&		&	Significant periods & & FWHM   	\\
                &	density	&	noise		&		&		& & 	\\
\kms     & 	Jy							& Jy				&					& Days      & Days & Days                    \\
\hline
-91.505	&	4.0 						& 0.9			& 5				& 507, 370, 255 & 507.5 & 19.3 \\
-90.802	&  6.5						& 0.9			& 8				& 507, 368, 255 & 507.6 & 15.8\\
-85.578	&	14.3						& 0.9			& 16				& 517, 254 & 512.8 & 28.1 \\
-84.920	&	11.6						& 0.9			& 13				& 816, 503 & 508.2 & 27.7\\ 
-84.305	& 22.0						& 0.9			& 23				& 512 & 511.9 & 28.5\\
\hline
\end{tabular}
\end{center}
\end{table*}

\subsection{G338.93-0.06}

This source does not appear to have been mapped at cm wavelengths. 

\begin{figure}
\resizebox{\hsize}{!}{\includegraphics[clip]{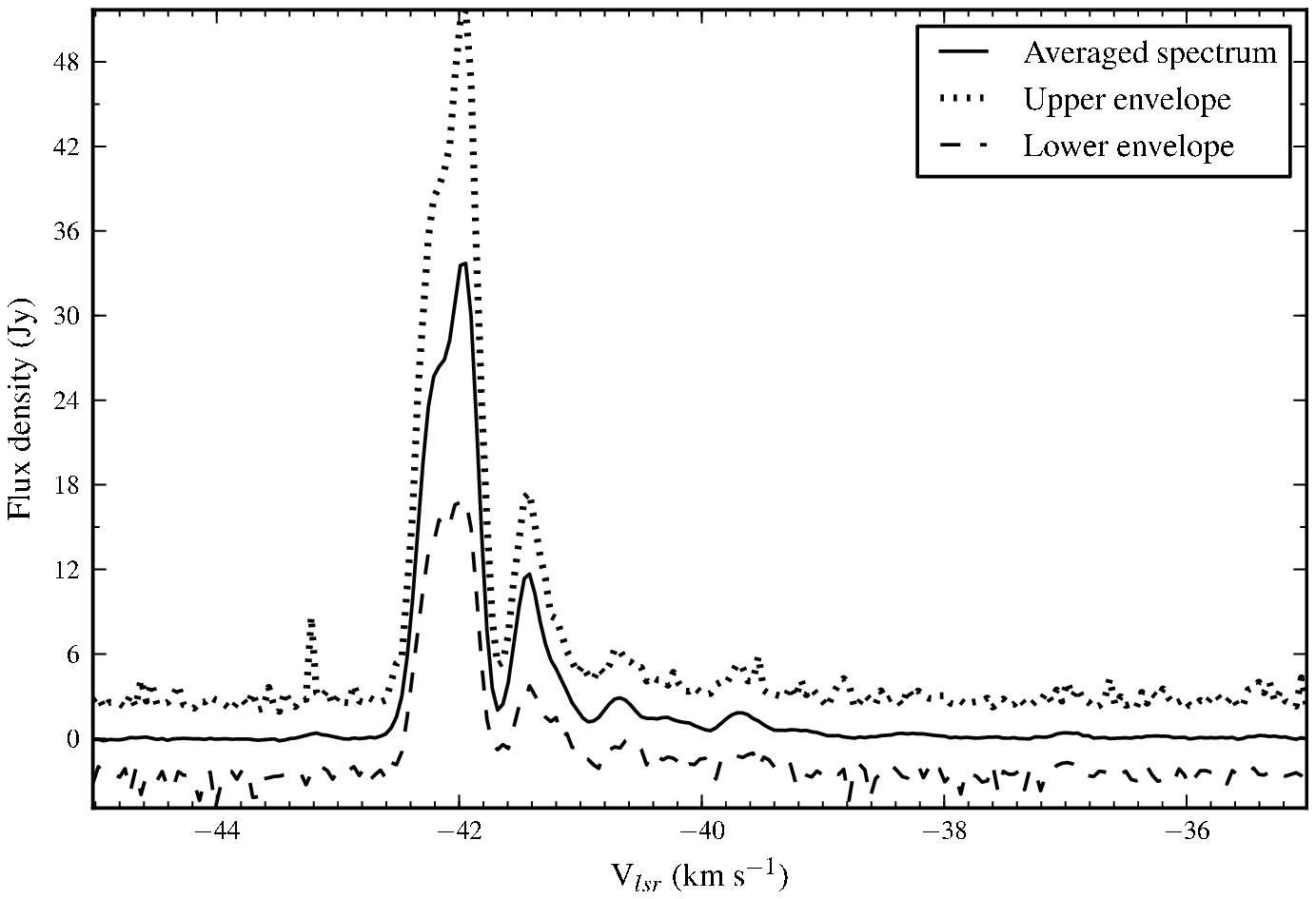}}
\caption{Range of variation across all spectral channels for G338.93-0.06 at 6.7 GHz during 2003--2008.}
\label{fig:g3389_67_spectra}
\end{figure}

\begin{figure}
\resizebox{\hsize}{!}{\includegraphics[clip]{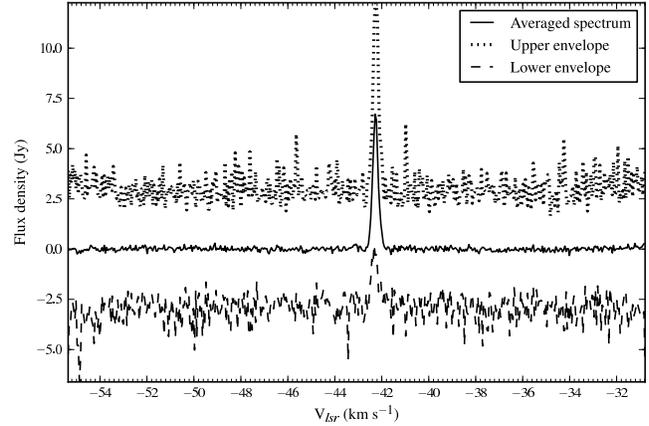}}
\caption{Range of variation across all spectral channels for G338.93-0.06 at 12.2 GHz.}
\label{fig:g3389_122_spectra}
\end{figure}

\begin{figure}
\resizebox{\hsize}{!}{\includegraphics[clip]{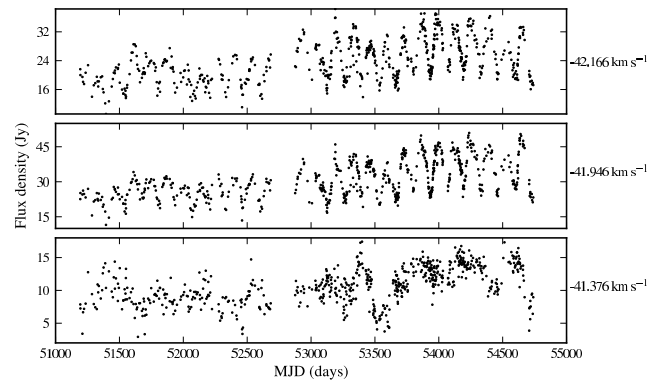}}
\caption{6.7 GHz time series for peak velocity channels in G338.93-0.06.}
\label{fig:g3389_67_ts}
\end{figure}

\begin{figure}
\resizebox{\hsize}{!}{\includegraphics[clip]{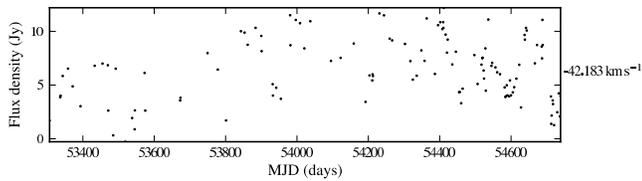}}
\caption{12.2 GHz time series for the peak velocity channel in G338.93-0.06.}
\label{fig:g3389_122_ts}
\end{figure}

Figures~\ref{fig:g3389_67_spectra} and \ref{fig:g3389_122_spectra} show the spectra at 6.7 and 12.2 GHz.  The time series at 6.7 and 12.2 GHz are shown in Figures~\ref{fig:g3389_67_ts} and \ref{fig:g3389_122_ts}.  Two of the peaks at 6.7 GHz show correlated periodic variability.  The third peak at -41.376 shows variability and also shows significant power in the Lomb-Scargle periodogram. The periods found are summarised in Table~\ref{tab:g3389-freqs}.  The 12.2 GHz peak shows the same periodicity as the 6.7 GHz peaks at -42.166 and -41.946 \kms but is not as well sampled as at 6.7 GHz.  The epochfolded periodograms also indicate a remarkable level of periodicity. We derive a weighted mean period using the longer time-series at 6.7 GHz of 132.8$\pm$0.8 days. The wave form of this source is different from the other periodic sources since it shows a very sharply defined minimum and does not have a quiescent period. No discernable delays were found between the two periodic features.

\begin{table*}
\caption{Periods from Lomb-Scargle periodogram and epoch-folding for  G338.93-0.06}
\label{tab:g3389-freqs}
\begin{center}
\begin{tabular}{lllllll}
\hline
Velocity & mean & mean & S/N & L-S & E-F period & E-F  \\
                & flux 	&	rms		&		&	Significant periods & & FWHM   	\\
                &	density	&	noise		&		&		& & 	\\
\kms     & 	Jy							& Jy				&					& Days      & Days & Days                    \\
\hline
\multicolumn{7}{c}{6.7 GHz}\\
-42.166	& 24.0						& 0.9			& 26				& 1158, 132.6, 87.4, 84.9, 66.4  & 132.8 & 1.4\\
-41.946	& 31.2						& 1.0			& 32				& 132.6, 66.4  & 132.9 & 1.1\\
-41.376	&	10.9						& 0.8			& 13				& 2837, 1320, 396.8, 297.1, 235.4, 201.9, 86.8 & -- & -- \\
\multicolumn{7}{c}{12.2 GHz}\\
-42.183	& 6.43						& 1.0			& 6				& 132.4	& 132.6 & 4.4 \\
\hline
\end{tabular}
\end{center}
\end{table*}

\subsection{G339.62-0.12}

\begin{figure}
\resizebox{\hsize}{!}{\includegraphics[clip]{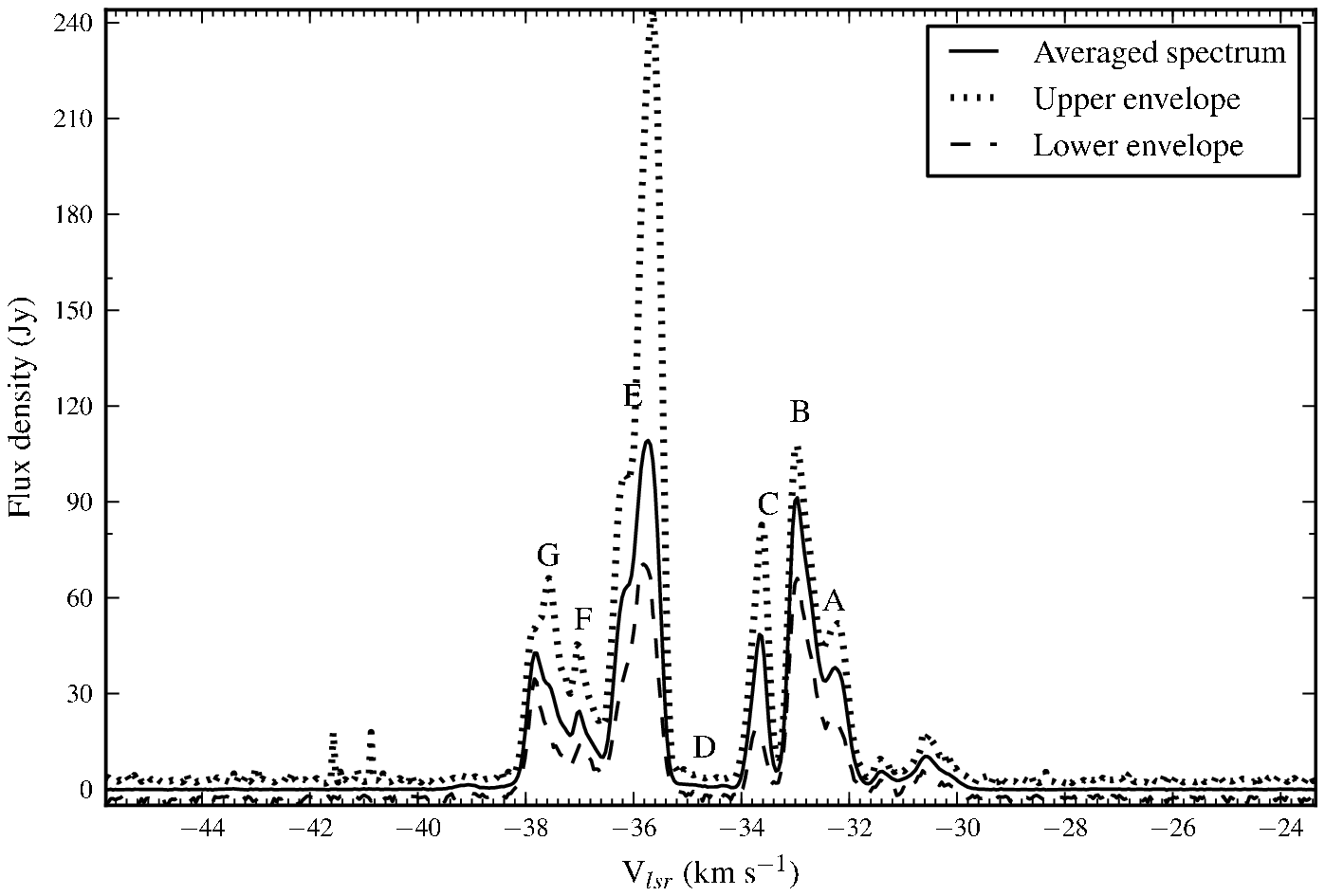}}
\caption{Range of variation across all spectral channels for G339.62-0.12 at 6.7 GHz during 2003--2008.}
\label{fig:g3396_67_spectra}
\end{figure}

\begin{figure}
\resizebox{\hsize}{!}{\includegraphics[clip]{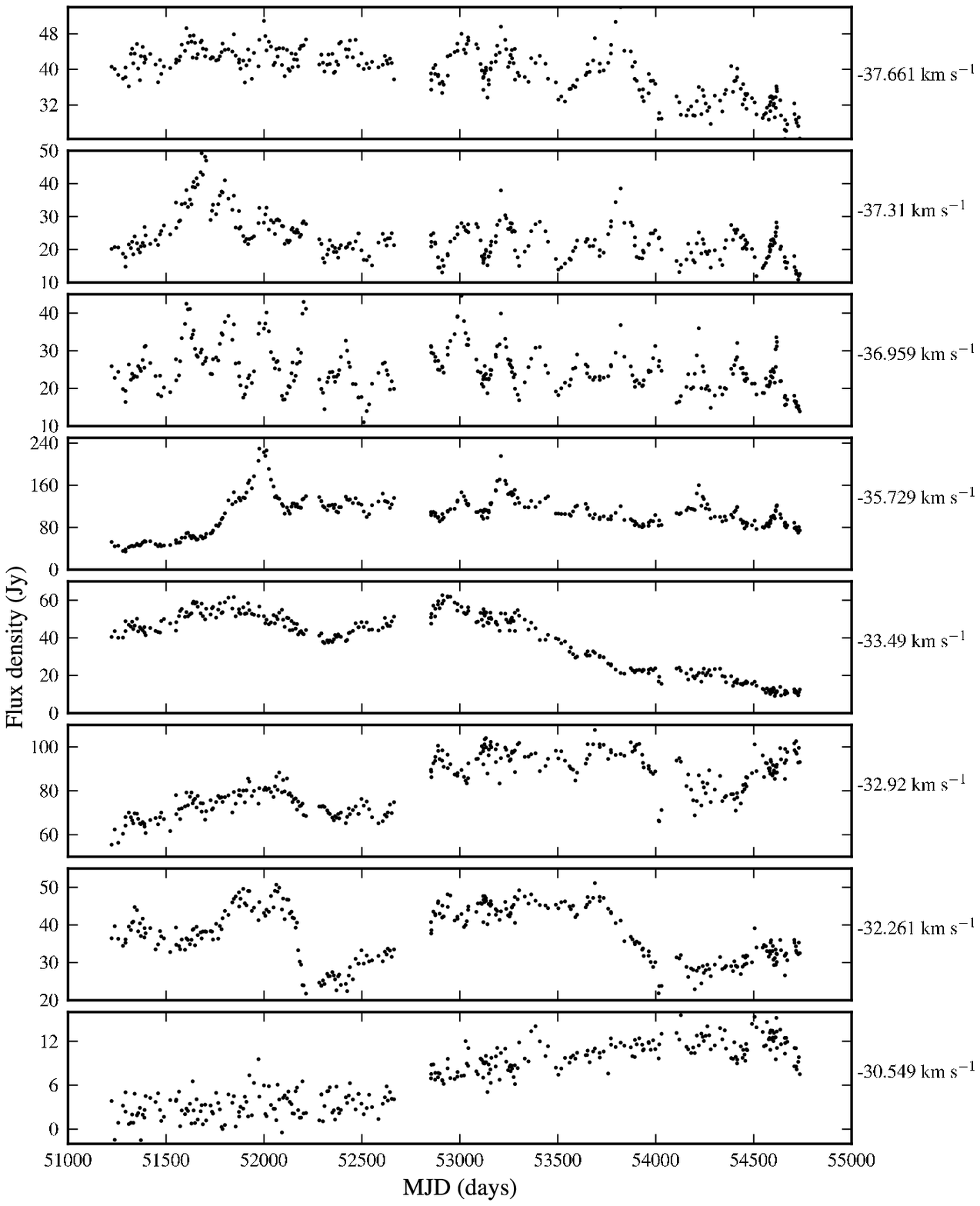}}
\caption{6.7 GHz time series for peak velocity channels in G339.62-0.12.}
\label{fig:g3396_67_ts}
\end{figure}

\begin{figure}
\resizebox{\hsize}{!}{\includegraphics[clip]{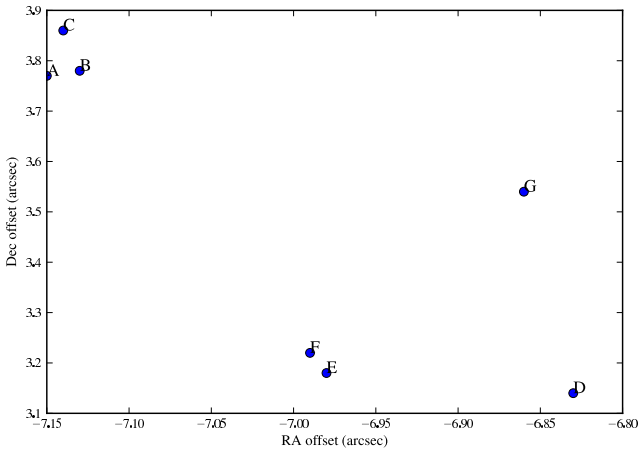}}
\caption{Distribution of maser spots for G339.62-0.12 from table 2 of \citet{Walsh1998}}
\label{fig:g3396_spotmap}
\end{figure}

The 6.7 GHz spectra are shown in Figure~\ref{fig:g3396_67_spectra} with maser features mapped by \citet{Walsh1998} indicated.  The time-series for the peak velocity channels is shown in Figure~\ref{fig:g3396_67_ts}.  The features at -37.661, -37.310, -36.959, and -32.92 \kms show correlated periodic variations. We plotted the  maser spots listed in table 2 of \citet{Walsh1998}  in Figure~\ref{fig:g3396_spotmap} since it is necessary to understand the variability of the different features. The feature at -30.549 \kms was not detected by \citet{Walsh1998} but has been steadily increasing in intensity.   Table ~\ref{tab:g3396_67-freqs} summarises the peaks in the Lomb-Scargle and epoch-folding periodograms.  Each individual feature shows its own uncorrelated variability, in addition to the shared periodic flares. Spots A, B and C form a tight cluster to the north-east of the region and spot D seems to have vanished. Spots E, F and G are slightly dispersed but all show a common periodicity. The same periodicity is probably also present in spot B.     We adopt a  weighted-mean period of 200.3$\pm$1.1 days. The uncorrelated variability as well as the destruction and formation of new maser spots seems to point to volatile local conditions.   No discernable delays were found. 

\begin{table*}
\caption{Periods from Lomb-Scargle periodogram and epoch-folding for G339.62-0.12 at 6.7 GHz}
\label{tab:g3396_67-freqs}
\begin{center}
\begin{tabular}{lllllll}
\hline
Velocity & mean & mean & S/N & L-S & E-F period & E-F  \\
                & flux 	&	rms		&		&	Significant periods & & HWHM   	\\
                &	density	&	noise		&		&		& & 	\\
\kms     & 	Jy							& Jy				&					& Days      & Days & Days                    \\
\hline
-37.661	& 39.3						& 1.2			& 32				& 685.5, 201 & 200.9 & 6.5	\\
-37.310	& 23.4						& 1.1			& 22				& 1606, 1222, 892, 703, 557, 199 & 198.7 & 4.6 \\
-36.959	& 25.1						& 1.1			& 23				& 1249, 200 & 200.6 & 2.7 \\
-35.729	& 106.4						& 2.0			& 54				& 2677, 1196, 446, 201 & 200.7 & 8.5 \\
-33.490	& 38.4						& 1.3			&	31				& 1479 & -- & --\\
-32.920	& 83.4						& 1.7			& 50				& 1653, 936, 202&  201.0 , 5.1 \\
-32.261	& 37.3						& 1.2			& 31				& 1653, 892 & -- & --\\
-30.549	& 7.3						& 1.0			& 7				& 2555  & -- & --\\
\hline
\end{tabular}
\end{center}
\end{table*}

\section{Discussion}

We summarise the confirmed periods in Table~\ref{tab:vital_stats}, including the number of observations, the time-span covered and the number of cycles observed.  The observations cover a sufficiently long time-span to confirm true periodicity in these sources. The range of periods observed (including the 29.5 day period reported for G12.89+0.49) can only be readily explained by orbital motions, most probably due to a binary system.

\begin{table}
\caption{Summary of periods and the observation statistics}
\label{tab:vital_stats}
\begin{center}
\begin{tabular}{lrrrrr}
\hline
Source(transition)				& Period & $\sigma$P	& num.	& time span	& num.  \\
										& (days)		&		(days) &			obs			&(days	)		& cycles\\
\hline
G9.62+0.19E (6.7 GHz)		& 243.3 & 2.1		& 708						& 3544			& 14.5	\\
G9.62+0.19E (12.2 GHz)		& 	& 	& 617						& 3111			& 12.7	\\
G188.95+0.89 (6.7 GHz)	& 395 & 8	& 488						& 3544			& 9.1 \\
G188.95+0.89 (12.2 GHz) & 	&	& 373						& 3102			& 7.9 \\
G328.24-0.55 (6.7 GHz)	& 220.5 & 1.0		& 320						& 3483			& 15.8 \\
G328.24-0.55 (12.2 GHz) & &	&	112						& 1432			& 6.5	\\
G331.13-0.24 (6.7 GHz)	& 509 & 10	& 318						& 3519			& 6.9 \\
G338.93-0.06 (6.7 GHz)	& 132.8 & 0.8 	& 653						& 3544			& 26.7 \\
G338.93-0.06 (12.2 GHz)	& 	& 	& 123						& 1431			& 10.7 \\
G339.62-0.12 (6.7 GHz)	& 200.3 & 1.1		& 315						& 3513			& 17.5 \\
\hline
\end{tabular}
\end{center}
\end{table}

\begin{figure}
\resizebox{\hsize}{!}{\includegraphics[clip]{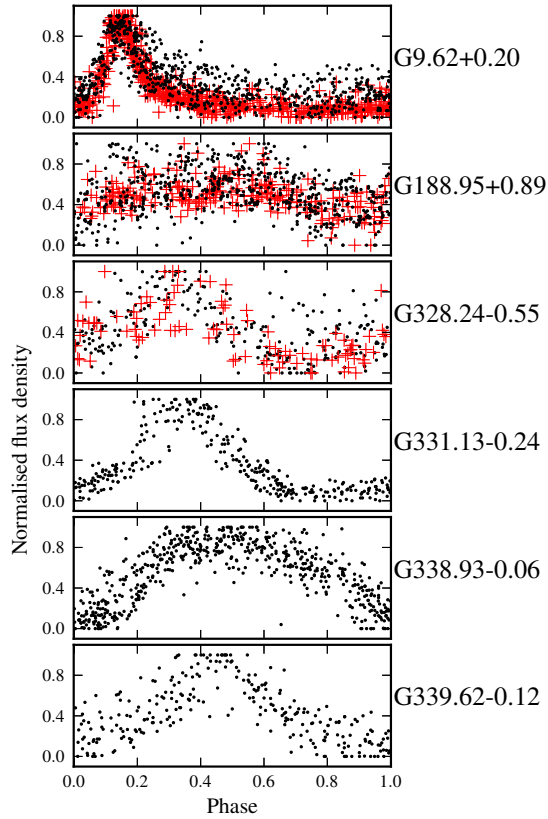}}
\caption{Normalised folded waveforms for each periodic source. The dots are for 6.7 GHz measurements and the crosses are for 12.2 GHz, where available.} 
\label{fig:norm_fold}
\end{figure}

Figure~\ref{fig:norm_fold} shows a comparative diagram of the normalised folded wave forms for a representative feature from each source.  The normalised fold was produced by dividing the time-series into cycles based on its best-fit period, and the flux density measurements in each cycle were divided by the maximum recorded in the cycle.  The time-axis was normalised by dividing by the period. Some generalisations can be made regarding the waveforms.  G9.62+0.20, G328.24-0.55 and G339.62-0.12 show sharply-peaked asymmetric pulse profiles, while G188.95+0.89 is closer to sinusoidal. The simple colliding wind binary model of \citet{VanderWalt2011} is able to reproduce these waveforms by modification of the orbital parameters of the binary system. 

The behaviour of  G338.93-0.06, on the other hand, with very sharply-defined minima and no quiescent phase, cannot be explained by the colliding wind binary model. In the model, changes in the free-free emission amplified by the maser is due to a combination of the orbital motion of the secondary star and the partial recombination of the ionized gas at the ionization front. The peak of the maser flare corresponds to periastron passage while the minimum corresponds to apastron. Under the assumption of the adiabatic cooling of the shocked gas, the luminosity of shocked gas varies as $1/r$ where $r$ is the distance between the two stars. The luminosity and therefore the flux of ionizing photons increases rapidly as periastron is approached followed by a slower decay due to the recombination of the partially ionized gas in the ionization front for a rather eccentric orbit. This explains the asymmetric flare profile of some of the maser sources such as seen in, for example, G9.62+0.20E.  However, the $1/r$ dependence will always result in a rather slow increase of the flux of ionizing photons just after apastron. It is therefore difficult to explain the sharp increase in the maser flux density as in the case of G338.93-0.06 within the framework of the colliding-wind binary model.

G331.13-0.24 with its changes in pulse profile between cycles and features is also not consistent with a simple colliding wind binary model. We note that G12.89+0.49, while having the shortest period known, also has similar characteristics to G331.13-0.24, with a very stable phase for the minima of the light curve but varying phase for the peak.
G331.13-0.24 is the only source in this sample associated with a known outflow.   It is possible that the methanol masers, while not directly associated with the outflow, could be amplifying emission from an episodic or precessing outflow.  One of the mechanisms put forward by \citet{MacLeod1996} for strong phase lags in methanol maser flares in G351.78-0.54 was an intermittent thermal jet. A strong bipolar outflow has subsequently been found towards this source \citep{Leurini2009}.

There is also a very close correspondence between 12.2 and 6.7 GHz waveforms, where we are able to achieve sufficient signal to noise, indicating a common mechanism.  \citet{VanderWalt2009} have also observed simultaneous flaring at 107 GHz for G9.62+0.19E. The 12.2 GHz masers show higher amplitude variations, in general.

The periods appear to be stable for all of the sources.  The spectral structure of the sources also appears to be stable in general, although some features may slowly change in intensity over time. The masers tend to return to a similar intensity after a flare, indicating that the maser region itself is not affected by the mechanism causing the periodic modulation.  This was confirmed by high resolution observations of G9.62+0.20E during a flare \citep{Goedhart2005}.  However, the varying amplitudes of flares and long term trends may be due to local changes in maser path length.  

It may be significant that all of these sources have both 6.7 and 12.2 GHz methanol masers, even if they may not be strong enough for monitoring with the 26m telescope.  All of the sources, with the exception of G188.95+0.89, appear to have OH maser emission as well.  The statistical analysis of \citet{Breen2011a} indicates that these objects are at a more advanced evolutionary stage than those exhibiting only 6.7 GHz maser emission. Most of the periodic sources, with the exception of G188.95+0.89 and possibly G339.62-0.12 have been getting brighter.

Understanding the underlying mechanism of these periodic variations is limited by observational constraints.  Monitoring of the flux density of the associated HII regions is challenging due to confusion and the variable uv-coverage of interferometric arrays. However, monitoring of other maser species may enable us to constrain whether the variability lies with the seed or pump photons. Mainline hydroxyl masers are believed to have  a common pump mechanism with class II methanol masers and simultaneous monitoring of methanol and hydroxyl masers will be valuable. \cite{Green2012} looked for variations in the OH maser in G12.89+0.89 and found an indication that there may be a drop in intensity coinciding with the same effect at 6.7 GHz.  A programme to monitor the OH masers using the HartRAO 26m and the newly commissioned KAT-7 telescope has been started and should give us greater insight into the variability of OH masers associated with periodic methanol sources.  \citet{Araya2010} find quasi-periodic variations in the formaldehyde maser in IRAS18566+0408, with correlated variations in some of the 6.7 GHz methanol maser components. The pump mechanism of formaldehyde is not fully understood -  \cite{Boland1981} show that the masers can be pumped by the free-free continuum radiation from an HII region, while  \citet{Araya2010} argue that the simultaneous flaring of methanol and formaldehyde indicate a common infrared pump and that the variability is caused by periodic accretion of circumbinary disk material.

\section{Conclusions}

We have presented 10 years of monitoring of six periodic class II methanol masers.  The periods in this sample range from 132.8 to 509 days.  The regularity of the flares indicate a periodic underlying mechanism, while the amplitude of the maser response can vary.  Where it is possible to monitor the 12.2 GHz methanol masers, these have been found to flare simultaneously with their 6.7 GHz counterparts.

While the cause of the periodicity is yet to be confirmed, it seems likely that these sources are associated with binary systems. Further work needs to be done in developing time-dependent maser pump models and in finding associated variability in other tracers.

\section*{Acknowledgements}
Part of S Goedhart's work on this project has been supported by the National Research Foundation under grant number 74886.

\bibliographystyle{mn2e}
\bibliography{./mn-jour,./periodicity}

\label{lastpage}

\end{document}